\documentclass{llncs}
\usepackage{graphicx}
\usepackage[a4paper, top=2.54cm, bottom=2.54cm, left=2.54cm, right=2.54cm]{geometry}
\usepackage[T1]{fontenc}
\usepackage{enumitem}
\usepackage{amsmath, amssymb}

\usepackage{amsthm}
\usepackage{faktor} 
\usepackage{multirow}
\usepackage{hyperref}
\usepackage{url}
\usepackage{booktabs}
\usepackage{subfig}
\usepackage{xcolor}

\let\llncssubparagraph\subparagraph
\let\subparagraph\paragraph

\usepackage{titlesec}

\let\subparagraph\llncssubparagraph

\usepackage{graphicx}
\graphicspath{ {./figure/} }
\usepackage{adjustbox}

\usepackage{algorithm}
\usepackage{algpseudocode}

\newcommand{\hnet}{CaRENets}

\newcommand{\ahemat}{A\(^\ast\)HEMAT}

\newcommand{\sealver}{v2.3.1}

\begin{document}

\date{}

\title{\hnet: Compact and Resource-Efficient CNN for Homomorphic Inference on Encrypted Medical Images}

%

\author{
Jin Chao\inst{1} \and
Ahmad Al Badawi\inst{1} \and
Balagopal Unnikrishnan\inst{2} \and
Jie Lin\inst{1} \and
Chan Fook Mun\inst{1} \and
James M. Brown\inst{3} \and
J. Peter Campbell\inst{4} \and
Michael Chiang\inst{4} \and
Jayashree Kalpathy-Cramer\inst{3} \and
Vijay Ramaseshan Chandrasekhar\inst{1} \and
Pavitra Krishnaswamy\inst{1} \and
Khin Mi Mi Aung\inst{1} \\~\\
\{Jin\_Chao, Ahmad\_Al\_Badawi, Mi\_Mi\_Aung\}@i2r.a-star.edu.sg}


\institute{Institute for Infocomm Research, A*STAR, Singapore \and
National University of Singapore \and
Massachusetts General Hospital, Harvard Medical School, MA, US \and
Oregon Health \& Sciences University, Portland, OR, US
}
\maketitle
\begin{abstract}
Convolutional neural networks (CNNs) have enabled significant performance leaps in medical image classification tasks. However, translating neural network models for clinical applications remains challenging due to data privacy issues. Fully Homomorphic Encryption (FHE) has the potential to address this challenge as it enables the use of CNNs on encrypted images. However, current HE technology poses immense computational and memory overheads, particularly for high resolution images such as those seen in the clinical context.

We present CaRENets: \textbf{C}ompact and \textbf{R}esource-\textbf{E}fficient CNNs for high performance and resource-efficient inference on high-resolution encrypted images in practical applications.
At the core, CaRE\-Nets comprises a new FHE compact packing scheme that is tightly integrated with CNN functions.
CaRENets offers dual advantages of memory efficiency (due to compact packing of images and CNN activations) and inference speed (due to reduction in number of ciphertexts created and the associated mathematical operations) over standard interleaved packing schemes.

We apply \hnet~to perform homomorphic abnormality detection with 80-bit security level in two clinical conditions - Retinopathy of Prematurity (ROP) and Diabetic Retinopathy (DR). The ROP dataset comprises 96 $\times$ 96 grayscale images, while the DR dataset comprises 256 $\times$ 256 RGB images. We demonstrate over 45x improvement in memory efficiency and 4-5x speedup in inference over the interleaved packing schemes. As our approach enables memory-efficient low-latency HE inference without imposing additional communication burden, it has implications for practical and secure deep learning inference in clinical imaging.
\end{abstract}

\section{Introduction}\label{sec:Introduction}
Convolutional neural networks (CNNs) have enabled significant performance leaps in medical image classification tasks. To leverage these advances in clinical practice, several stakeholders are working towards cloud-based health services that can receive patient data, apply previously trained neural network models for inference, and return resulting predictions to the referral source. Although cloud-based inference obviates the need for expensive high-performance hardware at point of care, it involves sharing of sensitive patient data and poses important privacy concerns. Much of the recent work on privacy-preserving deep learning has been focused on developing strategies for training CNNs across multiple institutions without sharing protected health information \cite{mcmahan2016communication,chang2018distributed,ryffel2018generic}. We instead consider prevention of unauthorized breach of patient confidentiality by encrypting the data before transmitting to the cloud. In this scenario, trained CNN models are deployed in cloud infrastructure and applied to encrypted data for inference. 

Fully Homomorphic Encryption (FHE) caters to the need for computing on encrypted data, and has been applied to computations involving CNNs on image datasets ranging from MNIST~\cite{MSFT:DGL+16,badawi2018alexnet} and CIFAR-10~\cite{hesamifard2017cryptodl} to medical images~\cite{chou2018faster}. However, homomorphic CNN approaches do not easily scale to the image resolution, resource efficiency and latency requirements in typical medical image classification tasks. Most studies using CNNs for medical image classification consider images with resolutions ranging 96 $\times$ 96 to 512 $\times$ 512~\cite{greenspan2016guest}. Current FHE methods are impractical for such tasks, as they require large amount of memory (e.g., 789 GB$\sim$1183 GB) for mid-range resolution medical images~\cite{chou2018faster}. As such, homomorphic CNN evaluation for practical medical image classification requires strategies to dramatically improve the computational and memory efficiency.

Here, we present a novel resource-efficient method for homomorphic CNN inference on encrypted medical images. Our approach overcomes the need for large cloud memory, yields good latency, and does not impose additional network communication burden.

To motivate specifications for our approach, we consider two ophthalmology application scenarios involving classification of Retinopathy of Prematurity (ROP) and Diabetic Retinopathy (DR). 
\begin{enumerate}
\item The first application involves ROP screening in low resource settings. ROP is a leading cause of preventable blindness worldwide. As such, several efforts are underway, particularly in rural areas in low and middle income countries, to screen babies for this condition using portable cameras including some with mobile phones. Quantitative screening tools could reduce dependence on expert graders \cite{li2012telemedicine,richter2009telemedicine}. Recent work has demonstrated that CNNs are effective for automatic detection of ROP with fundus photos \cite{brown2018automated}. A cloud-based service for ROP classification with CNNs would benefit from the ability to preserve privacy. At the minimum, such applications would require 96 $\times$ 96 images, and latencies within 30 minutes.  
\item The second application involves fundus testing for DR as part of multi-center clinical trials. These applications employ clinical-grade equipment in disparate sites and pool the data for integrated analyses. Several recent studies have demonstrated the use of CNNs for automatically classifying severity of DR \cite{gulshan2016development}. Clinical trials providers would benefit from the ability to perform automated DR classification while also fulfilling their ethical commitments to maintain patient privacy. At the minimum, such applications would require 256 $\times$ 256 images and latencies within a few hours.  
\end{enumerate}
We therefore consider the problem of resource-efficient homomorphic CNN inference for ROP and DR images of resolution 96 $\times$ 96 - 256 $\times$ 256.  We assume a secure machine learning service paradigm wherein (a) suitable trained CNN models are available for the use case of interest, and (b) the data referral source has the ability to encrypt images before initiating a request for cloud-based inference. Such approaches are particularly desirable for retinal images as they carry a surprising amount of information about overall patient health \cite{poplin2018prediction}, and also biometric identification information due to unique vascular patterning \cite{jain2004introduction}. 

The major contributions of our paper are summarized as follows.
\begin{itemize}
    \item We propose a new compact packing strategy to pack matrices and vectors into HE-encrypted ciphertexts, and build a new matrix computation library, \ahemat, that leverages the SIMD-like execution model associated with modern FHE schemes to facilitate general computations on encrypted matrices such as: addition, multiplication and transposition.
    
    \item We further design and develop a compact and resource-efficient homomorphic CNN, \hnet, which is built on the tight integration of the optimized matrix-vector multiplication function of \ahemat~and the CNN layers. We show that \hnet~are capable of addressing a variety of image classification problems, ranging from standard benchmarking datasets to real-world medical imaging datasets.
    
    \item We provide a set of experiments, detailed analysis and comparisons with state-of-the-art solutions to evaluate the performance of our packing technique and \hnet, and demonstrate \hnet~can improve both memory efficiency and inference speed significantly.
\end{itemize}

\textbf{Paper Organization.}
Section~\ref{section:Preliminaries} introduces background and related work. Section~\ref{sec:hemat} describes A*HE\-MAT, our technique to pack matrices into ciphertexts and do the computations on matrices. Section~\ref{sec:carenets} details our methods and optimizations to develop~\hnet. We conduct performance evaluation of \hnet~with MNIST benchmark in Section~\ref{sec:hnet:performance:evaluation}. In Section~\ref{sec:real:world:usecases}, we introduce two real-world use cases for medical image inference, and evaluate the performance of \hnet~on the associated image datasets. Finally, in Section~\ref{sec:discussion} and Section~\ref{sec:conclusion} we discuss our findings, draw some conclusions and outline directions for future work.

\section{Background and Related Work}\label{section:Preliminaries}
In this section, we provide an overview of the core FHE and deep learning concepts used in this work.

\subsection{Fully Homomorphic Encryption}
FHE is an emerging technology that enables computing on encrypted data in the absence of the decryption key. The idea was born in 1978~\cite{FOSC:RivAdlDer78} and materialized into a secure theoretical scheme in 2009~\cite{STOC:Gentry09} using the bootstrapping technique that allows ciphertext refreshing. Basically, we start by a noisy encryption scheme that masks the secret message by some noise that can be completely filtered out by the decryption procedure. Moreover, the scheme includes a budget for computation that allows performing a certain number of operations on ciphertexts without decryption. The computation budget decreases as we perform more operations due to the noise growth. Bootstrapping is used to reduce the noise and produce a substitute ciphertext with lower noise and higher computation budget. Unfortunately, bootstrapping is generally too costly for practical applications despite the massive improvements that have been proposed in the HE literature so far. Therefore, HE practitioners opt to use a more efficient variant of FHE known as levelled-FHE~\cite{brakerski2014leveled} that can be configured to support computations with a certain multiplicative depth without bootstrapping.

FHE has gone through a sequence of improvements in functionality and performance~\cite{van2010fully,brakerski2014leveled,fan2012somewhat,chillotti2016faster,cheon2017homomorphic,cheonhomomorphic} that enabled the development of interesting privacy-preserving applications~\cite{graepel2012ml,lauter2014private,cheon2015homomorphic,xie2016privlogit,bonte2018privacy}. In this study, we use the Fan-Vercauteren (FV)~\cite{EPRINT:FanVer12} levelled-FHE that is implemented in Microsoft SEAL~\cite{SEAL} in all our experiments. 

In most of current FHE schemes, the plaintext space can be factorized into multiple spaces using the Chinese Remainder Theorem~\cite{DCC:SmaVer14}. This allows one to pack a vector of plaintext messages and operate on them in a SIMD manner. Adding (resp. multiplying) two packed ciphertexts results in component-wise addition (resp. multiplication) of the vectors. Moreover, mature implementations of FHE schemes provide mechanisms to rotate the underlying plaintext vectors inside the ciphertexts to enable inter-slot interaction. For instance, to add two numbers in misaligned slots, one needs to rotate one vector to align the corresponding slots followed by component-wise addition. Moreover, one can extract desired values by multiplying by a mask vector of bits. It should be noted that packing is done at the data encoding phase of the computation right before encryption. 

\subsection{Deep Learning}
Modern Deep Neural Networks, such as Convolutional Neural Networks (CNN), Generative Adversarial Networks (GAN), Long Short-Term Memory networks (LSTM), are composed of linear and nonlinear function blocks. Linear function blocks like Convolution, Full-connected, Shortcut, Element-wise addition, Batch Normalization, Average Pooling, are basically matrix multiplications with simple multiplication and addition operators. Nonlinear function blocks usually contain more complex operations such as exponential in Softmax/Sigmoid or max operation in Max pooling/ReLU layer. The depth of neural networks is increased by repeating the linear and nonlinear blocks in the network architecture. With increasing scale and difficulty of target applications (e.g. from MNIST to CIFAR-10 to ImageNet), the model capacity of neural networks has to be enlarged (e.g. wider and deeper networks) in order to meet the desired prediction accuracies.

Applying homomorphic encryption (HE) in deep neural networks poses unique challenges. 
The HE algorithms today only support addition-multiplication operations and are extremely slow, often taking tens of minutes to process a single low resolution image given a shallow network.
Several specific changes have to be made to the neural network algorithms for it to be compatible with HE algorithms. For instance, reducing numerical precision of network weights to save bitwidth for polynomial coefficients, approximating nonlinear operations (e.g. Max, ReLU, Sigmoid, etc) with low-degree polynomial functions, limiting input size and network width/depth to deal with resource-hungry HE algorithms given devices with limited computational and memory capacity. All these changes would potentially lead to a considerable drop in prediction accuracy.

\subsection{Related Work}
The literature is rich with studies dealing with privacy-preserving prediction as a service. We are more interested in solutions that involve computing on encrypted data using FHE. These related works are introduced in the following paragraphs.

CryptoNets~\cite{MSFT:DGL+16} was the first work that showed how to evaluate the inference phase of CNNs on encrypted images. The authors used a simple 5-layer network to predict the likelihood values of the MNIST~\cite{MNIST} dataset (28 $\times$ 28 $\times$ 1) with high accuracy level (99\%) despite the employment of low-precision training, approximate activation functions and shallow network structure. Of particular interest is the packing scheme used in CryptoNets which enabled the prediction of multiple images in a single network evaluation. In CryptoNets, ciphertexts contain 8192 slots that can be used to pack 8192 different pixels. In ciphertext $i$, the authors pack pixel $i$ from 8192 images. Since each image has 784 pixels, they can pack 8192 images in 784 ciphertexts. Running the network on these ciphertexts will generate the predictions for all images simultaneously. This packing scheme is ideal for high-throughput scenarios where the user requests the prediction of a large bundle of images. On the other hand, single-image prediction takes the same time and resources. Hence, this packing scheme is not ideal for single-image (especially for high-resolution images) prediction scenarios as it requires a large number of ciphertexts. Hereafter, we refer to this packing scheme as interleaved packing.

A major limitation of the interleaved packing is the requirement for enormous computing resources. For instance, Faster CryptoNets~\cite{chou2018faster} used a somewhat similar packing scheme (but only for a single image) as CryptoNets and showed how to run deeper networks on high-resolution medical images. For inference of one medical image of resolution (224 $\times$ 224 $\times$ 3), Faster CryptoNets requires 1183.8 GB RAM on a high-end computing platform ( n1-megamem-96). 

More related to our work is E2DM~\cite{jiang2018secure} which exploits the SIMD-like execution model in modern FHE schemes to facilitate general matrix computations such as addition, multiplication and transposition. As a use case, the authors could homomorphically evaluate a CNN to classify the MNIST dataset. Although E2DM provides efficient computational complexity, it suffers from the following limitations: 1) E2DM assumes the matrices are square and that the ciphertext is large enough to contain the entire matrix, both are not usually the case in real-world applications. 2) This adds extra burden to the user to compute on large matrices as it poses a need to manually decompose large matrices into smaller blocks. 3) More importantly, for deep learning problems, E2DM requires the user to know some information about the model such as filter and stride sizes, which can impose privacy concerns if the model is confidential. 

To resolve the limitations above, we propose a new packing scheme. Our packing scheme is compact as it utilizes the number of available slots in ciphertexts. Hence, the proposed scheme is resource-efficient in terms of memory requirement. Moreover, it is flexible and user-friendly as it can be naturally used to perform computation on matrices with arbitrary sizes without zero-padding. Although our scheme requires a relatively larger number of rotations on the ciphertext slots, our experiments demonstrate that this extra cost can be amortized by the SIMD processing our packing scheme exploits.

\section{\ahemat - General Computation library for Compactly Packed Matrices}\label{sec:hemat}

One of the key mathematical computations in neural networks is the matrix (including vector) multiplication. To enable easy efficient integration of neural networks and homomorphic encryption, we have designed and implemented the \ahemat~library, which can encrypt matrices and vectors into ciphertexts and perform matrix computations such as addition, multiplication and transposition in the encrypted domain. Generally, \ahemat~encrypts a matrix of $m$ rows and $n$ columns into one of the following four layouts:
\begin{enumerate}
\item Row Packing (RP): each row of the matrix is stored into the slots of a separate ciphertext. Thus the encrypted matrix is composed of $m$ ciphertexts.

\item Column Packing (CP): each column of the matrix is stored into the slots of a separate ciphertext. Thus the encrypted matrix is composed of $n$ ciphertexts.

\item Row Compact Packing (RCP): The matrix is stored into the ciphertext slots consecutively in the row-major order. The encrypted matrix is composed of $k$ ciphertexts where $k = \lceil\frac{m \cdot n}{s}\rceil$ and $s$ is the number of slots in ciphertext.

\item Column Compact Packing (CCP): The matrix is stored into the ciphertext slots consecutively in the column-major order. Similar to RCP, the encrypted matrix is composed of $k$ ciphertexts.
\end{enumerate}
\ahemat~provides functions to convert an encrypted matrix between any of the four layouts. For instance, matrix multiplication can be done most efficiently if the operands are: 1) a RP/RCP matrix and 2) a CP/CCP matrix. Algorithm~\ref{alg:mat:mul} shows the multiplication of two $d \times d$ square matrices $A$ and $B$, where $A$ is in RCP layout whereas $B$ is in CCP layout. For simplicity, we show the case where the number of ciphertexts required to pack $A$ or $B$ is $k = 1$. Nevertheless, \ahemat~ is flexible to support cases where $k > 1$.

	 \begin{algorithm}[!ht]
	 \scriptsize
	     \caption{Matrix multiplication via \ahemat~using RCP by CCP packing layouts.} \label{alg:mat:mul}
	     \begin{algorithmic}[1]
	         \Statex{Let $d$ denote the dimension of the input square matrices $A$ and $B$. We assume that $A$ and $B$ are packed in RCP and CCP layouts, respectively. Let $(\bar{\cdot})$ denote the ciphertext encrypting the quantity ($\cdot$).}
	         \vspace{.5em}
	         \Statex{\textbf{Input}: $\bar{A}$, $\bar{B}$}
	         \Statex{\textbf{Output}: $\bar{C} = \bar{A} \cdot \bar{B}$}
	         \vspace{.5em}
	         \State{$\bar{C}$ = Enc(0)}
	         \For{$i = 1~\text{to}~d$ }
	             \State{$\bar{P} = \text{Mult}(\bar{A}, \bar{B})$}
	             \State{$\bar{S} = \text{PartialSum}(\bar{P}, d)$}\Comment{Summation for every $d$ slots of $\bar{P}$ into $\bar{S}$ slots with $log_2{}d$ rotations}
	             \For{$j = 1~\text{to}~d$ }
	                 \State{$\bar{t} = \text{CMult}(\bar{S}, \text{maskVector})$} \Comment{multiply with one-hot maskVector to get only one slot of $\bar{S}$ }
	                 \State{$\bar{t} = \text{rot\_row}(\bar{t}, \text{diff})$} \Comment{Rotate the current slot in $\bar{S}$ to get aligned with the final result}
	                 \State{$\bar{C} = \bar{C} + \bar{t}$}
	             \EndFor
	             \State{$\bar{B} = \text{rot\_row}(\bar{B}, d)$}
	         \EndFor
	         \State \textbf{return} $\bar{C}$
	     \end{algorithmic}
	 \end{algorithm}

Algorithm~\ref{alg:mat:mul} requires $O(d)$ Mult (ciphertext-ciphertext multiplication) and $O(d^2)$ CMult (ciphertext-constant multiplication), as well as $O(d^2)$ rotations. The consumed multiplication depth is 1 Mult $+$ 1 CMult. Note that, among all these operations, Mult is the most computationally intensive in terms of runtime and noise growth. 

Besides the encrypted matrix computation, \ahemat~also provides the multiplication function of plaintext matrix with ciphertext matrix to enable privacy-preserving Machine Learning as a Service (MLaaS) on encrypted images (provided by the user) using a pre-learned model owned by the cloud or a service provider. The implementation details of this particular scenario are provided in the subsequent section.

\section{CaRENets}\label{sec:carenets}
By leveraging the packing technique and matrix-vector multiplication function in \ahemat, we design and implement a compact and resource-efficient homomorphic CNN, called \hnet. 
In this section, we describe how the matrix-vector multiplication function is implemented and how \hnet is built by integrating the function with a convolutional neural network. For \hnet, We encrypt the input image compactly into ciphertexts whereas the weight vectors are plaintext. \hnet is designed to homomorphically evaluate the inference phase of MLaaS on encrypted images.

\subsection{Image Packing}
In our packing scheme, the pixels of one image are flattened into a big vector, which is subsequently mapped into one or multiple ciphertexts. To make our packing scheme compatible with the traditional CNNs, we have redesigned the procedures of convolutional layers and fully connected layers to use only matrix operations. As a result, we need a few to hold the input or output values for each CNN layer, and the total amount of memory usage of \hnet~is significantly minimized. Furthermore, by minimizing the number of ciphertexts and the associated addition and multiplication operations, the processing time of one round of prediction, aka, the latency for predicting one single image, is also reduced.
\begin{figure*}[!ht]
    \centering
    \includegraphics[width=.85\textwidth, height=5.5cm]{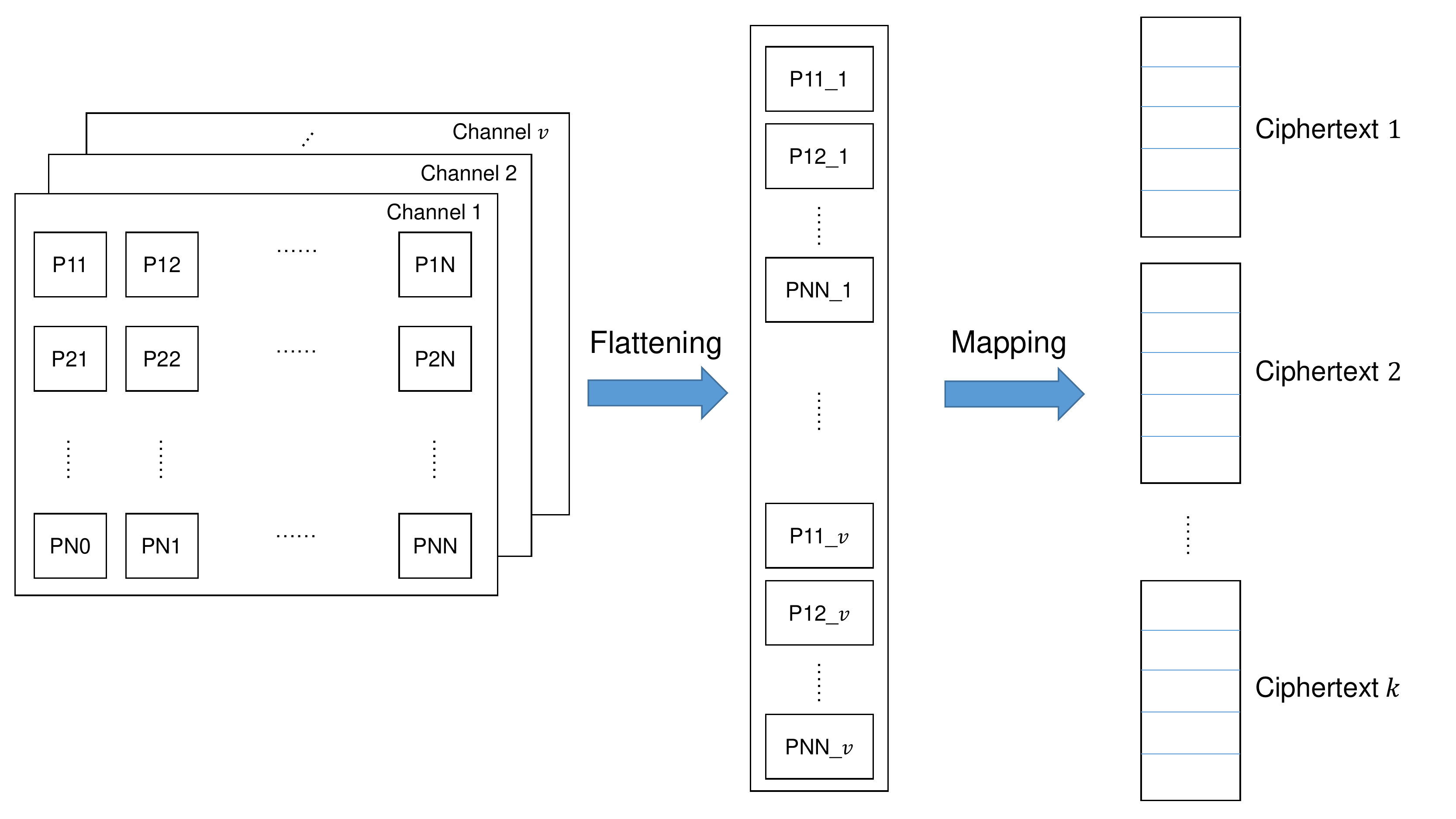}
    \vspace{-0.3cm}
    \caption{Mapping the pixels of the input image to the ciphertext slots.}
    \label{fig:flattened:image}
\end{figure*}

As seen in Figure~\ref{fig:flattened:image}, an image is composed of a 2-D array of pixels, with each pixel having multiple channels such as Red, Green, and Blue. Therefore, each image can be digitized into a 3-dimensional (3-D) tensor. Next, the 3-D tensor is flattened into a 1-D vector according to a certain sequence, e.g., channel by channel. We  note that, in the whole processing flow of \hnet, the inputs and outputs of each layer are always represented as one virtual 1-D vector. 

Each virtual 1-D vector is mapped into one or multiple ciphertexts. In other words, the values in the vector are stored in the slots of the ciphertexts contiguously. \hnet~uses the parameter \textsf{slots\_used}, which can be configured by the user, to control how many slots in each ciphertext will be used to store the values. When one ciphertext is filled with the maximum amount of values, it will switch to the next ciphertext.

\subsection{Design of \hnet~Layers}
One cannot directly use the traditional deep learning frameworks (e.g., TensorFlow) when running CNN models with homomorphic encryption. Instead, a customized framework built solely on FHE primitives, i.e., additions and multiplications with ciphertexts, is needed. More specifically, we design the convolutional layer and fully connected layer with matrix-based operations, which are represented by the products of the weight matrices and the virtual 1-D vector storing all the inputs of the layer. 

\begin{figure*}[!ht]
    \centering
    \includegraphics[width=.75\textwidth, height=5.5cm]{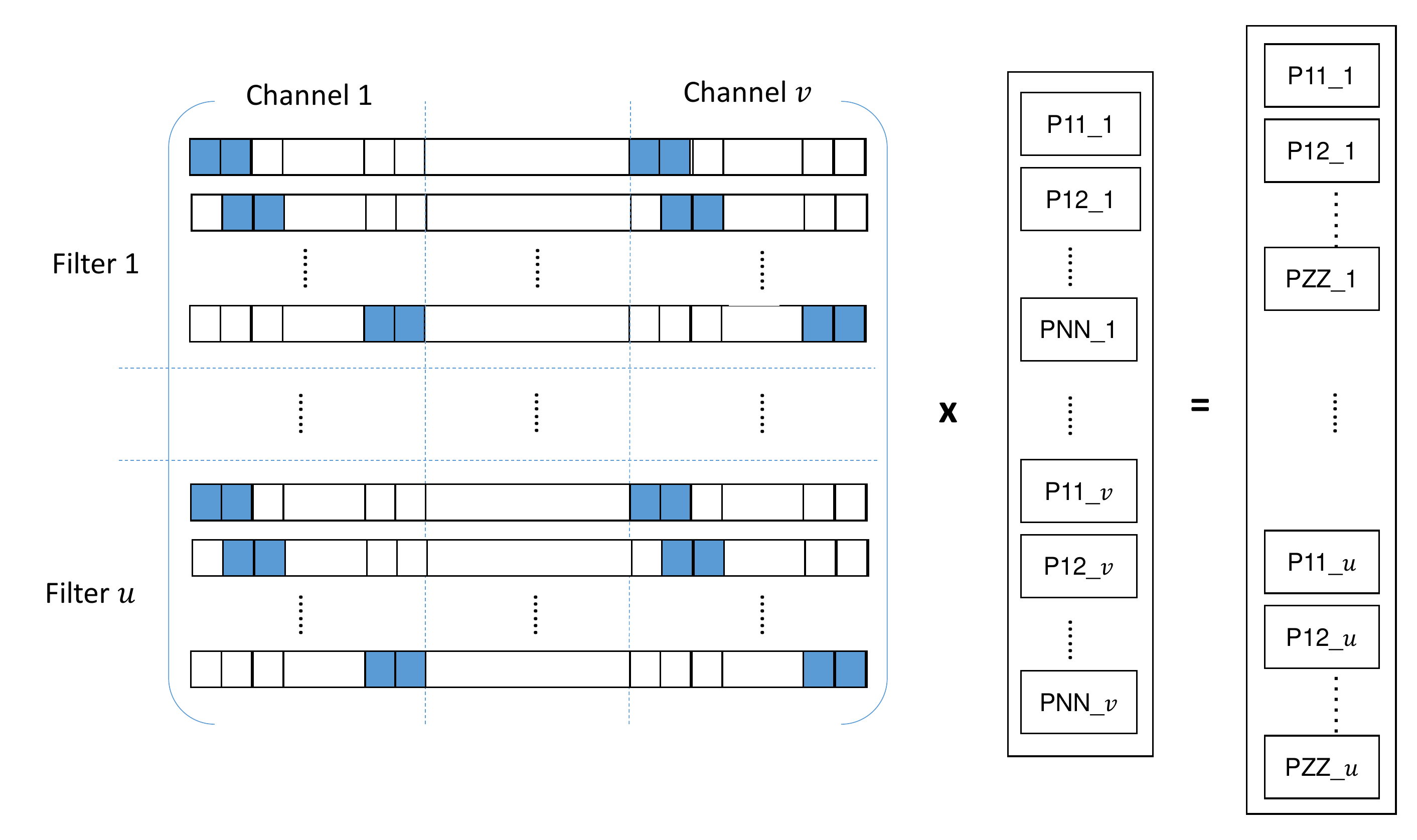}
    \vspace{-0.3cm}
    \caption{Design of the vector matrix product in the convolution layer in\hnet. The quantity ($ZZ$) represents the size of the output of the convolution layer, computed as a function of the image dimensions, padding and stride size.}
    \label{fig:plain:weight:ctxt:vector:mul}
\end{figure*}

The dot product of one row in the weight matrix and the virtual 1-D vector produces exact one element in the output 1-D vector. A bias can also be added into the result if defined by the model. For the convolutional layer, the weight matrix is composed of all the possible shifted locations of all the filters as shown in Figure~\ref{fig:plain:weight:ctxt:vector:mul}. In each row, the slots where the filter is located are filled with the weights, while all the rest slots are padded with zero. For the fully connected layer, the procedure is similar, except that the number of rows in the weight matrix is equal to the number of output neurons, and the weights in each row corresponds to the input neurons of the fully connected layer. 

\subsection{Parallel Processing}
In Figure~\ref{fig:plain:weight:ctxt:vector:mul}, the dot products of the rows in the weight matrix and the 1-D vector holding all the inputs composes the output 1-D vector. All the rows can be processed independently, thus it can be naturally paralleled. In our implementation, we use \textsf{OpenMP} to parallel the multiplication of the weight matrix and input vector for each convolutional or fully connected layer.

\begin{figure*}[!ht]
    \centering
    \includegraphics[width=.95\textwidth, height=5cm]{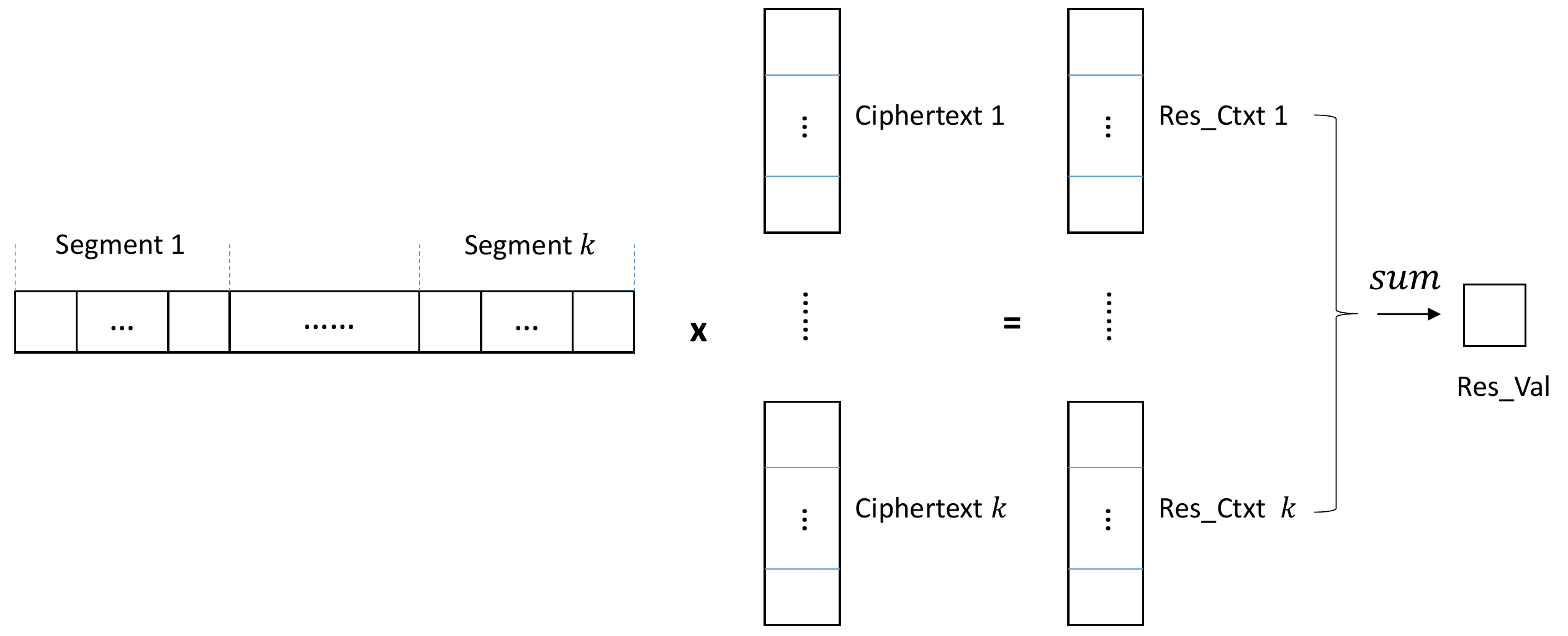}
    \caption{Parallel processing using \textsf{OpenMP} for improved performance on multi-core platforms.}
    \label{fig:parallel:processing}
\end{figure*}

There is another level of parallelism which is inside the multiplication of one row with the 1-D input vector. Recall that the 1-D vector is actually mapped into $k$ separate ciphertexts. The weight row can be further divided into $k$ segments, with each segment multiplying with one of the $k$ ciphertexts as shown in Figure~\ref{fig:parallel:processing}. After getting the $k$ intermediate resultant ciphertexts, we first add them together to get the summed ciphertext, and then perform Halevi and Soup's \textsf{All Sum}~\cite{halevi2014algorithms} to add all the slots in the summed ciphertext, which produces the dot product of the entire row with the input vector.

There might be all-zero segments in the weight row. When encountering an all-zero segment, we just skip this segment and the corresponding ciphertext. This optimization removes the unnecessary multiplications in the process.
\subsection{FHE-Friendly Deep Learning}
The following are the major considerations to generate a deep learning model that is FHE-Friendly.
\begin{enumerate}
\item Simplified CNN Architecture: For CNNs, a major obstacle for translation to the homomorphic domain is the activation functions since they are usually not polynomials, and therefore unsuitable for evaluation with FHE schemes. We use a simplified network structure with square activation function as shown in Table~\ref{tab:network:architecture}.
\begin{table*}[htbp]
\scriptsize
    \centering
    \renewcommand{\arraystretch}{1.3}
    \caption{Architecture of \hnet~for both training and inference.}
    \begin{tabular}{l p{7.5cm}}
        \toprule
        \multicolumn{1}{c}{\text{ Layer Type}} & \multicolumn{1}{c}{\text{ Description}} \\ \midrule
        \multirow{2}{*}{Convolution} & \multirow{2}{7.5cm}{\(25\) filters of size \(5 \times 5\) and stride \((2,2)\) without padding.} \\ 
        & \\ \hline
        \multirow{1}{*}{Square} & \multirow{1}{*}{Outputs of the previous layer are squared.} \\ \hline
        \multirow{2}{*}{Convolution} & \multirow{2}{7.5cm}{\(50\) filters of size \(5 \times 5\) and stride \((2,2)\) without padding.} \\ 
        & \\ \hline
        \multirow{1}{*}{Square} & \multirow{1}{*}{Outputs of the previous layer are squared.} \\ \hline
        \multirow{3}{*}{Fully Connected} & \multirow{3}{7.5cm}{Weighted sum of the entire previous layer with \(10\) filters, each output corresponding to \(1\) of the possible \(10\) digits.} \\ 
         & \\
         & \\ \hline
    \end{tabular}
    \label{tab:network:architecture}
\end{table*}

\item Low-Precision Training: As most of FHE scheme schemes deal only with integer arithmetic we scale the normalized input images and weights to integers by multiplying them by 2- to 4-bit scale factors (integers) and round them to integers similar to fixed-point encoding.

\item Choice of Parameters: FHE parameters need to be configured cautiously to facilitate homomorphic computation, ensure a minimum security level and achieve optimal performance. We follow the scheme used in HCNN~\cite{badawi2018alexnet} to configure the FHE parameters.
\end{enumerate}

\section{MNIST Benchmarking}\label{sec:hnet:performance:evaluation}
We use the standard MNIST benchmark dataset to evaluate the performance of our packing scheme and \hnet. MNIST comprises of 70,000 28 $\times$ 28 grayscale images of handwritten digits (Arabic numerals 0-9). We train a FHE amenable CNN on the standard 60,000 training split, and test on the remaining 10,000 images. 

\textbf{Hardware Configuration}: All our experiments were performed on a server with an Intel\(^\text{\textregistered}\) Xeon\(^\text{\textregistered}\) Platinum 8170 CPU @ 2.10 GHz with 26 cores, \(188\) GB RAM.

\textbf{Efficiency of our Packing Scheme}: We compare performance with the interleaved packing used in CryptoNets~\cite{MSFT:DGL+16}. To ensure fair comparison, we implement the network shown in Table~\ref{tab:network:architecture} using the two packing schemes: 1) interleaved packing and 2) our compact packing. Both implementations are done in SEAL \sealver~ and executed on the same machine. The system parameters used in this experiment and performance results are shown in Table~\ref{tab:mnist:performance}. It can be clearly seen that our compact packing scheme is more efficient than the interleaved packing in terms of runtime (5.1$\times$) and memory usage (5.9$\times$). Although our packing scheme requires higher multiplicative depth due to the rotation operations, we can still achieve better performance due to the low number of ciphertexts used. 
\vspace{-0.5cm}
\begin{table}[htbp]
\scriptsize
  \centering
  \caption{Performance on MNIST dataset using interleaved and compact packing schemes for a single image inference. Time, memory and security level $lambda$ units are: sec, GB and bits, respectively.}
    \begin{tabular}{l c c c r c c}
    \toprule
    Packing scheme~~~ & {$\log_2{}n$} & ~~~{$\log_2{}q$} & ~{$t$} & ~~Time & ~~Memory & ~~~$\lambda$ \\
    \midrule
    Interleaved~\cite{MSFT:DGL+16} & 14    & 400   & 4398047232001 & 155.4 & 12.1 & > 80  \\
    Compact & 14  & 540  & 4398047232001 & 30.1 & 2.06 & > 80 \\
    \bottomrule
    \end{tabular}%
  \label{tab:mnist:performance}%
\end{table}%

\section{Real-World Use-Case Evaluations}\label{sec:real:world:usecases}

We demonstrate the proposed homomorphic evaluation packing strategy on two medical image datasets and evaluate the results for classification accuracy, memory efficiency, and latency. The two datasets have distinct characteristics to showcase generalizability of our method. While the ROP data constitutes posterior pole images segmented for vasculature, the DR data comprises color RGB retinal fundus images. 

\textbf{Retinopathy of Prematurity (ROP) Data}: We obtained 1,000 posterior pole retinal RGB photographs collected as part of the ongoing Imaging and Informatics in ROP (i-ROP) study~\cite{{brown2018automated}}. Each image was assessed for the ROP characteristics of retinal arterial tortuosity and venous dilation at the posterior pole~\cite{capone2006standard}, and denoted as normal or plus by at least three independent experts. A reference standard diagnosis (RSD) label was assigned to each image based on adjudication among three experts, resulting in 884 normal and 116 plus images. We converted the RGB images to grayscale and applied gamma adjustment and contrast-limited adaptive histogram equalization (CLAHE). Because plus disease predominantly affects the retinal vasculature, we segmented the vessels \cite{brown2018automated}, and downsampled the resulting images to 96 $\times$ 96 for classification experiments.

\textbf{Diabetic Retinopathy (DR) Data}: We obtained 249 color retinal fundus images from the IDRiD challenge dataset collected at an eye clinic located in Nanded, Maharashtra, India~\cite{porwal2018indian}. Each image is annotated for features indicating the presence of DR: microaneurysms, hemorrhages, exudates, venous beading, microvascular abnormality, and neovascularization in retinal fundus photographs~\cite{gulshan2016development}, and denoted as healthy or disease. The dataset had 168 healthy and 81 disease cases. We resized the original images to 256 $\times$ 256 and normalized them by the maximum intensity value in each image.  
\begin{figure*}[!ht]
    \centering
    \includegraphics[width=.75\textwidth, height=3cm]{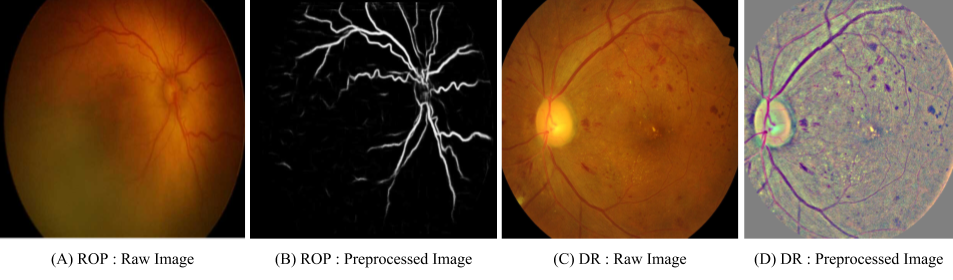}
    \vspace{-0.3cm}
    \caption{Exemplar illustrations of the raw and pre-processed images for the ROP and DR clinical  use cases}
    \label{fig:retinal:images}
\end{figure*}

\subsection{Training procedures} 
We randomly split the pre-processed datasets into training, validation and testing cohorts with split ratios 7:1:2 for ROP images, and 3:1:1 for DR images. Before training, we augmented the training dataset using a combination of flipping and 90 degree rotations, and randomly sampled the augmented dataset to balance the distribution of classes. For each case, we trained the FHE model and a comparable supervised baseline. The splitting, training, and augmentation procedures were identical for both the FHE model and supervised baselines.  We tuned the FHE model by varying the dropout rates and number of filters until there was negligible effect on the validation loss. 

\subsection{Accuracy }
In each case, we repeated training with 4 different random seeds, and report mean classification AUCs with associated standard deviations. The results are reported in Table 3, and show that the FHE model achieves comparable classification performance (albeit with $\approx 3 \%$  drop in AUC) to the supervised baselines.
\begin{table}[htbp]
\scriptsize
	\centering
	\caption{Classification Results: Clinical Image Use Cases}

	\begin{tabular}{l l r r }
	\toprule
		Dataset ~~~~~~~~~~~~~~~~~~&Models~~~~~~~~~~~~~~& AUC~~~~~~~~~ & Accuracy 	\\
	\midrule
	\multirow{2}{*}{ROP 96 $\times$ 96} 
		&HE-Model~~~~~~~~~ & 0.888 $\pm$ 0.034 ~~~~~~~~~ & 0.925 $\pm$ 0.006 	\\
		&Inception-V3~~~~~~~~~ & 0.915 $\pm$ 0.022 ~~~~~~~~~ & 0.905 $\pm$ 0.024 \\
	\midrule
	\multirow{2}{*}{IDRID 256 $\times$ 256}
	    &HE-Model ~~~~~~~~~& 0.799 $\pm$ 0.065 ~~~~~~~~~& 0.725 $\pm$ 0.062 \\
        &CIFAR-Net ~~~~~~~~~& 0.829 $\pm$ 0.054 ~~~~~~~~~& 0.765 $\pm$ 0.068 \\
	\bottomrule
	\end{tabular}
\end{table}%

\subsection{Performance Evaluation}
\textbf{\hnet~Performance on ROP}: We evaluate the performance of \hnet~on the ROP dataset. The same 5-layer network shown in Table~\ref{tab:network:architecture} is used with both packing schemes. Table~\ref{tab:rop:performance} shows the system parameters and performance results. Compared with the MNIST results, the results of this experiment demonstrate larger gains with our \hnet~with compact packing over the interleaved packing. This suggests that the benefits of \hnet~are even larger as we handle higher resolution images. More specifically, compact packing enables a 4$\times$ improvement over interleaved packing in runtime and a 45.9$\times$ improvement in memory. These results are not counter-intuitive since an image in the ROP dataset is 96 $\times$ 96 $\times$ 1 and requires a large number of ciphertexts to run the 5-layer network using the interleaved packing.
\begin{table}[htbp]
\scriptsize
  \centering
  \caption{Performance on ROP dataset using interleaved and compact packing schemes for a single image inference. Time, memory and security level $lambda$ units are: sec, GB and bits, respectively.}
    \begin{tabular}{l c c c r c c}
    \toprule
    Packing scheme~~~ & {$\log_2{}n$} & ~~~{$\log_2{}q$} & ~{$t$} & ~~Time & ~~Memory & ~~~$\lambda$ \\
    \midrule
    Interleaved~\cite{MSFT:DGL+16} & 14    & 450   & 4503599627763713 & 3946.4 & 135 & > 80  \\
    Compact & 14  & 660  & 4503599627763713 & 994.9 & 2.94 & > 80 \\
    \bottomrule
    \end{tabular}%
  \label{tab:rop:performance}%
\end{table}%

\textbf{\hnet~Performance on IDRiD}: In this last experiment, we target a more challenging problem and run \hnet~on the IDRiD dataset, which has images of resolution 256 $\times$ 256 $\times$ 3 pixels. As shown in Table~\ref{tab:idrid:performance}, the memory resources (188 GB) on our test machine were not sufficient to run the 5-layer network using the interleaved packing. On the other hand, we could run the network with the compact packing scheme with evaluation time of less than 2 hours using 17.2 GB memory.

\begin{table}[htbp]
\scriptsize
  \centering
  \caption{Performance on IdRID dataset using interleaved and compact packing schemes for a single image inference. Time, memory and security level $lambda$ units are: sec, GB and bits, respectively.}
    \begin{tabular}{l c c c r c c}
    \toprule
    Packing scheme~~~ & {$\log_2{}n$} & ~~~{$\log_2{}q$} & ~{$t$} & ~~Time & ~~Memory & ~~~$\lambda$ \\
    \midrule
    Interleaved~\cite{MSFT:DGL+16} & 14    & 450   & 4503599627763713 & {$-$} & {> 188} & > 80  \\
    Compact & 14  & 660  & 4503599627763713 & 6004.7 & 17.2 & > 80 \\
    \bottomrule
    \end{tabular}%
  \label{tab:idrid:performance}%
\end{table}%

\section{Discussions}\label{sec:discussion}
Medical imaging applications require the ability to compute on encrypted data, in the absence of a decryption key. Homomorphic encryption provides significant security guarantees and ability to compute, but is bottlenecked by the high overhead of CNN computations, especially for high-resolution images. We have described a new resource efficient homomorphic encryption strategy, rooted in novel FHE compact packing strategies, to enable such computations with CNNs for medical image classification tasks. 

\textbf{Tradeoffs}: While we illustrated our method for two use cases, the approach is generalizable to other scenarios with due consideration of tradeoffs between resolution, accuracy, security and performance. FHE technology requires simplification of activation functions, network architectures and network depth. Such simplified networks have limited capacity for accurate classification of very high resolution images. We have outlined some possible tuning strategies to improve accuracy with the limited capacity network in these settings, but these will need to be considered on a case-by-case basis. 

\textbf{Possible Use Cases}: Given that our current methods lose about 3-5\% accuracy, we envision that the first use cases of homomorphic CNN inference will lie primarily in screening and triage applications. This is especially relevant in emergent scenarios where speedy privacy preserving computations would offer significant gains and aid the clinical workflows. 

\textbf{Extensions}: Our work focused on CPU implementations. Future work will enhance \hnet~methodologies for practical deployment by scaling to GPU implementations. For instance, HCNN~\cite{badawi2018alexnet} reports 50.51$\times$ improvement on GPU over CryptoNets~\cite{MSFT:DGL+16} on CPU in terms of inference speed. We foresee that our packing strategies can be similarly expanded to GPUs. Future studies will also address the message passing and computational resource aspects to scale with increasing resolution and security level requirements. 

\section{Conclusions}\label{sec:conclusion}
In this work, we first presented \ahemat, a library using a new compact packing strategy to encrypt matrices and vectors with homomorphic encryption, and provides general computation functions like additions, multiplications, transpositions on the encrypted matrices. On the basis of \ahemat, we further presented \hnet, a compact and resource-efficient CNNs for homomorphic inference on encrypted images. \hnet~leverages the compact packing technique and optimized matrix-vector multiplication function in \ahemat, and tightly integrates them into the CNN layers. We evaluated \hnet~ on the standard MNIST benchmark and showed significant gains in both memory efficiency and inference speed over the commonly used interleaved packing. We further evaluated \hnet~on abnormality detection with high-resolution retinal images. Across experiments, our results showed $4-5\times$ improvement over state-of-the-art in latency and $45-46\times$ gain in memory efficiency.  

\section{Acknowledgments}
This project was supported by funding from the Deep Learning 2.0 program at the Institute for Infocomm Research (I$^2$R), A*STAR, Singapore; 
research grants from the US National Institutes of Health (NIH grants R01EY19474, P30EY010572, and K12EY027720) and the US National Science Foundation (NSF grants SCH-1622679 and SCH-1622542); unrestricted departmental funding from the Oregon Health Sciences University, and a Career Development Award from  Research to Prevent Blindness (New  York, NY).


\bibliographystyle{splncs04}
\bibliography{biblio}

\end{document}